\documentclass{article}
\usepackage[margin=1in]{geometry}
\usepackage{amsmath}
\usepackage{setspace}
\usepackage{graphicx}
\usepackage{epstopdf}
\usepackage{booktabs}
\usepackage{threeparttable}

\title{Bootstrap Inference for Multiple Imputation under Uncongeniality and Misspecification}

\newcommand{\etal}{\textit{et al }}
\newcommand{\Var}{\mbox{Var}}

\author{Jonathan W. Bartlett \\ j.w.bartlett@bath.ac.uk \\ \vspace{0.1cm} \\ Rachael A. Hughes \\ Rachael.Hughes@bristol.ac.uk }

\begin{document}
\maketitle

\begin{abstract}
Multiple imputation has become one of the most popular approaches for handling missing data in statistical analyses. Part of this success is due to Rubin's simple combination rules. These give frequentist valid inferences when the imputation and analysis procedures are so called congenial and the complete data analysis is valid, but otherwise may not. Roughly speaking, congeniality corresponds to whether the imputation and analysis models make different assumptions about the data. In practice imputation and analysis procedures are often not congenial, such that tests may not have the correct size and confidence interval coverage deviates from the advertised level. We examine a number of recent proposals which combine bootstrapping with multiple imputation, and determine which are valid under uncongeniality and model misspecification. Imputation followed by bootstrapping generally does not result in valid variance estimates under uncongeniality or misspecification, whereas bootstrapping followed by imputation does. We recommend a particular computationally efficient variant of bootstrapping followed by imputation.
\end{abstract}

\section{Introduction}
\label{intro}
Multiple imputation (MI) has proven to be an extremely versatile and popular tool for handling missing data in statistical analyses. For a recent review, see \cite{murray2018}. Its popularity is due to a number of factors. The imputation and analysis stages are distinct, meaning it is possible for one entity to perform the imputation and another the analysis. It is flexible, in being able to accommodate various constraints and restrictions that the imputer or analyst may want to impose. Auxiliary variables can be used in the imputation process to reduce uncertainty about missing values or make the missing at random (MAR) assumption more plausible, yet need not be included in the analyst's model.

In MI the analysis model of interest is fitted to each imputed dataset. Estimates and standard errors from each of these fits are pooled using `Rubin's rules' \cite{Rubin:1987}. These give a point estimate as the simple average of the imputed data estimates. Rubin's variance estimator combines the average within imputation variance with the between imputation variance in estimates. This requires an estimator of the complete data variance, which for most estimators is available analytically.

In Rubin's original exposition the estimand was characteristic of a fixed finite population of which some units are randomly sampled and data are obtained \cite{Rubin:1987}. Rubin defined  conditions for an imputation procedure to be so called `proper' for a given complete data analysis. If in addition the complete data analysis gives frequentist valid inferences, MI using Rubin's rules yields valid frequentist inferences \cite{Rubin:1987,Rubin:1996,murray2018}. Subsequently Rubin's rules were criticised by some (e.g. \cite{fay1992inferences}) because in certain situations Rubin's variance estimator could be biased relative to the repeated sampling variance of the MI estimator. In response, Meng defined the concept of congeniality between an imputation procedure and an analyst's complete (and incomplete) analysis procedure \cite{Meng:1994}. If an imputation and analysis procedure are congenial, this implies the imputation is proper for the analysis procedure \cite{Nielsen2003}. Meng showed that for certain types of uncongeniality, Rubin's variance estimator is conservative, ensuring the intervals have at least the advertised coverage level \cite{Meng:1994}. In other settings however it can be biased downwards, leading to undercoverage of confidence intervals \cite{Robins/Wang:2000}.

Rubin's rules have proved fantastically useful since MI's inception, in particular because they facilitate the separation of imputation and analysis into two distinct parts and because they are so simple. Nevertheless, in settings where Rubin's variance estimator is asymptotically biased, if feasible, the analyst may desire sharp frequentist valid inferences. Robins and Wang proposed a variance estimator which is valid without requiring congeniality or correct model specification \cite{Robins/Wang:2000}. Their estimator requires calculation of various quantities depending on the estimating equations corresponding to the particular choice of imputation and analysis models. As such it is arguably harder to apply their approach when the imputer and analyst are separate entities. As far as we are aware, its use has been extremely limited thus far in practice due to these requirements.

Combining bootstrapping with MI was first suggested over 20 years ago \cite{shao1996bootstrap} and recently a number of papers have investigated a wider variety of approaches to combining them. Schomaker and Heumann investigated four variants which combined bootstrapping with multiple imputation \cite{schomaker2018bootstrap}. Their motivation for exploration of using bootstrap with MI was for situations where an analytical complete data variance estimator is not available, or one is concerned that the MI estimator is not normally distributed. On the basis of theoretical and empirical investigation, they recommended three of the four variants for use. They did not explicitly seek to investigate performance under uncongeniality or model misspecification however. von Hippel and Bartlett proposed an alternative combination of bootstrapping with MI in the context of proposing frequentist type (improper) multiple imputation algorithms, and noted that it would be expected to valid under uncongeniality \cite{vonHippelbartlett2019}. Lastly, Brand \etal investigated six different combinations of MI with bootstrapping in the context of handling skewed data, and recommended using percentile bootstrap confidence intervals with single (stochastic) imputation \cite{brand2019combining}.

In this paper we investigate the properties of the different combinations of MI and bootstrap which have been recommended by these previous papers, giving particular emphasis to their validity under uncongeniality. In Section \ref{mirubin} we review MI and Rubin's combination rules. In Section \ref{combiningBootMI} we describe the various combinations of bootstrapping and MI that have been recently recommended and consider their validity under uncongeniality. Section \ref{simulations} presents two sets of simulation studies, empirically demonstrating the impacts of uncongeniality on the frequentist performance of the different variants. We conclude in Section \ref{discussion} with a discussion.

\section{Multiple imputation using Rubin's rules}
\label{mirubin}
In this section we briefly review MI and Rubin's combination rules. In MI we first create $M$ imputed datasets. We fit our complete data model to each, obtaining estimates $\hat{\theta}_{m}$, $m=1,..,M$, and corresponding within imputation variance estimates $\widehat{\Var}(\hat{\theta}_{m})$. The estimate of $\theta$ is then given by $\hat{\theta}_{M}=M^{-1} \sum^{M}_{m=1} \hat{\theta}_{m}$, while the variance is estimated by
\begin{eqnarray}
\widehat{\Var}_{\text{Rubin}}(\hat{\theta}_{M}) = \left(1+\frac{1}{M} \right) \hat{\sigma}^{2}_{\text{btw}} + \hat{\sigma}^{2}_{\text{wtn}}
\label{rubinsVarEstimate}
\end{eqnarray}
where 
\begin{eqnarray*}
\hat{\sigma}^{2}_{\text{wtn}} = \frac{1}{M} \sum^{M}_{m=1} \widehat{\Var}(\hat{\theta}_{m})
\end{eqnarray*}
and
\begin{eqnarray*}
\hat{\sigma}^{2}_{\text{btw}} =  \frac{1}{M-1} \sum^{M}_{m=1} (\hat{\theta}_{m} - \hat{\theta}_{M})^{2}
\end{eqnarray*}

For subsequent developments, following von Hippel and Bartlett \cite{vonHippelbartlett2019} it will be useful to express each imputation estimate of $\theta$ as:
\begin{eqnarray*}
\hat{\theta}_{m} = \hat{\theta}_{\infty} + a_{m}
\end{eqnarray*}
where $\hat{\theta}_{\infty} = \lim_{M \rightarrow \infty} \hat{\theta}_{M}$, $\Var(\hat{\theta}_{\infty})=\sigma^{2}_{\infty}$, $E(a_{m})=0$ and $\Var(a_{m})=\sigma^{2}_{\text{btw}}$. Since the imputation estimates are conditionally independent given $\hat{\theta}_{\infty}$, we have that 
\begin{eqnarray}
\Var(\hat{\theta}_{M}) = \sigma^{2}_{\infty} + \frac{\sigma^{2}_{\text{btw}}}{M}
\label{varthetahatM}
\end{eqnarray}
Assuming the complete data analysis would provide valid frequentist inferences with complete data, if the imputation procedure is proper with respect to the complete data procedure \cite{Rubin:1987}, we have that
\begin{eqnarray*}
\sigma^{2}_{\infty} = \sigma^{2}_{\text{btw}} + \sigma^{2}_{\text{wtn}}
\end{eqnarray*}
so that 
\begin{eqnarray*}
\Var(\hat{\theta}_{M}) = \left(1+\frac{1}{M}\right)\sigma^{2}_{\text{btw}} + \sigma^{2}_{\text{wtn}}
\end{eqnarray*}
Inference for $\hat{\theta}_{M}$ is then performed assuming that
\begin{eqnarray*}
\frac{\hat{\theta}_{M}-\theta}{\sqrt{\widehat{\Var}_{\text{Rubin}}(\hat{\theta}_{M})}}
\end{eqnarray*}
is t-distributed with degrees of freedom given by
\begin{eqnarray*}
(M-1) \left \{ \frac{\hat{\sigma}^{2}_{\text{wtn}} + (1+M^{-1})\hat{\sigma}^{2}_{\text{btw}}}{(1+M^{-1})\hat{\sigma}^{2}_{\text{btw}}} \right \}^{2}
\end{eqnarray*}
These results were derived assuming that with complete data the degrees of freedom are infinite and $M$ is finite. In small sample settings the former assumption is questionable, and so Barnard \& Rubin subsequently proposed a small sample version of Rubin's rules \cite{Barnard/Rubin:1999}.

Meng subsequently defined an imputation procedure and a complete data analysis to be congenial essentially if there exists a Bayesian joint model for which the posterior distribution of the missing data matches that used by the imputation procedure and for which the complete data posterior mean and variance of the parameters of substantive interest are asymptotically the same as obtained by using the complete data analysis procedure \cite{Meng:1994}. Meng's congeniality definition in fact incorporated an additional notation of the analyst's incomplete data procedure, but for the present purposes this aspect is not relevant.

When the imputation and complete data analysis procedures models are congenial, this implies the imputation procedure is proper for the complete data analysis, and if in addition the complete data analysis gives frequentist valid inferences, Rubin's variance estimator for finite $M$ is asymptotically unbiased \cite{Nielsen2003}. When this is not case, Rubin's variance estimator can, depending on the configuration, be downwardly or upwardly biased as an estimator of the repeated sampling variance of $\hat{\theta}_{M}$ \cite{Meng:1994,Robins/Wang:2000}. 

Robins and Wang proposed a variance estimator for MI when each dataset is imputed using the maximum likelihood estimate of a parametric imputation model and the imputations are analysed using a non, semi or fully parametric model \cite{Robins/Wang:2000}. Their variance estimator is consistent without requiring the imputation and analysis models to be congenial nor even correctly specified. Hughes \etal compared Robins and Wang's proposal to Rubin's rules through a series of simulation studies where the imputation and analysis models were misspecified and/or uncongenial with each other \cite{hughes2016}. They demonstrated that Rubin's rules inference could be conservative or anti-conservative, whereas, at least for moderate or large sample sizes, inferences based on Robins and Wang's proposal were valid across their simulation scenarios. Hughes \etal noted however that a major practical obstacle to the widespread use of Robins and Wang's method is its implementation is specific to the particular imputation and analysis models, and no software currently implements it.

\section{Combining bootstrapping and multiple imputation}
\label{combiningBootMI}
In this section we review the combinations of bootstrapping and MI which have been recommended for use in the recent literature, and consider whether their validity in uncongenial settings.

\subsection{Imputation followed by bootstrapping}
The first collection of methods we consider are where MI is first applied, and then bootstrapping is applied to each imputed dataset. 

\subsubsection{MI boot Rubin}
The first combination considered (and recommended) by Schomaker and Heumann \cite{schomaker2018bootstrap} is standard MI using Rubin's rules, but using bootstrapping to estimate the within-imputation complete data variance:
\begin{enumerate}
	\item Impute the missing values in the observed data $M$ times, creating completed datasets $Y_{imp,m}$, $m=1,..,M$. Fit the analysis model to each, giving estimates $\hat{\theta}_{m}$.
	\item For each imputed dataset $Y_{imp,m}$, draw $B$ bootstrap samples with replacement
	\item For the $b$th bootstrap sample of the $m$th imputed dataset, estimate $\theta$, giving $\hat{\theta}_{m,b}$
	\item For imputation $m$, then calculate
	\begin{eqnarray*}
	\widehat{\Var}_{bs}(\hat{\theta}_{m}) = (B-1)^{-1} \sum^{B}_{b=1} (\hat{\theta}_{m,b} - \tilde{\theta}_{m})^{2}
	\end{eqnarray*}
	where $\tilde{\theta}_{m} = B^{-1} \sum^{B}_{b=1} \hat{\theta}_{m,b}$
	\item Rubin's rules is then applied with $\hat{\theta}_{m}$ ($m=1,..,M$) as the point estimates and $\widehat{\Var}_{bs}(\hat{\theta}_{m})$ ($m=1,..,M$) as the complete data variance estimates.
\end{enumerate}
This approach is what has often been used when no analytical estimator for the full data variance $\sigma^{2}_{\text{wtn}}$ is available, or if one is concerned about whether the analysis model is correctly specified. In the latter case, a sandwich variance estimator has sometimes been used to attempt to provide robustness to misspecification \cite{hughes2016}.

Provided the analysis model would give valid frequentist frequentist inferences with complete data, we expect MI boot Rubin to give asymptotically unbiased variance estimates of $\hat{\theta}_{M}$ under congeniality. This is supported by the setting 1 simulation results of Schomaker and Heumann \cite{schomaker2018bootstrap}. However, since this approach relies on Rubin's rules, we would not expect it to give unbiased variance estimates in general under uncongeniality. This hypothesis is supported by Schomaker and Heumann's setting 2 with high missingness simulation results, where we believe the imputation and analysis models are uncongenial, and where coverage for one parameter was $91\%$.

\subsubsection{MI boot pooled percentile}
The second approach considered and recommended by Schomaker and Heumann \cite{schomaker2018bootstrap} is the same as MI boot Rubin, except that Rubin's rules are not (directly at least) used:
\begin{enumerate}
	\item Impute the missing values in the observed data $M$ times, creating completed datasets $Y_{imp,m}$, $m=1,..,M$
	\item For each imputed dataset $Y_{imp,m}$, draw $B$ bootstrap samples with replacement
	\item For the $b$th bootstrap sample of the $m$th imputed dataset, estimate $\theta$, giving $\hat{\theta}_{m,b}$
	\item For point estimation of $\theta$, one can either use $\hat{\theta}_{M}$ or $(MB)^{-1} \sum^{M}_{m=1} \sum^{B}_{b=1} \hat{\theta}_{m,b}$.
	\item A $(1-2\alpha)\%$ percentile confidence interval for $\theta$ is formed by taking the $\alpha$ and $1-\alpha$ empirical percentiles of the pooled sample of $\hat{\theta}_{m,b}$ values
\end{enumerate}
This approach can be viewed as a route to obtaining a posterior credible interval, and hence assuming the analysis model would give valid inferences with complete data and that the imputation and analysis models are congenial, we expect it to give asymptotically unbiased variance estimates of $\hat{\theta}_{M}$. This is because first draws are taken from the posterior of the missing data given observed, and second, conditional on these, bootstrapping and estimating the parameters by their maximum likelihood estimate is in large samples equivalent to taking a draw from the posterior given the imputed missing data and the observed data \cite{Little/Rubin:2002}.

To explore this further, we can express the estimate from the $m$th imputation and $b$th bootstrap as
\begin{eqnarray*}
\hat{\theta}_{m,b} = \hat{\theta}_{\infty} + a_{m} + b_{mb}
\end{eqnarray*}
where $a_{m}$ is as defined earlier and $b_{mb}$ is a term representing the deviation due to bootstrap. The term $b_{mb}$ has mean zero and variance $\sigma^{2}_{\text{wtn}}$, since the between bootstrap variance for a completed dataset corresponds to the complete data variance. For $MB$ large, the sample variance of the pooled sample of $MB$ estimates, which we are effectively treating as a size $MB$ sample from the posterior, is:
	\begin{eqnarray}
	\Var_{MIBootPooled} = (MB)^{-1}  \sum^{M}_{m=1} \sum^{B}_{b=1} (\hat{\theta}_{m,b} - \hat{\theta}_{MB})^{2} \label{mibootvareq}
	\end{eqnarray}
where $\hat{\theta}_{MB}=(MB)^{-1} \sum^{M}_{m=1} \sum^{B}_{b=1} \hat{\theta}_{m,b}$. From standard results for the one-way random intercepts model \cite{searle2009variance}, this is an unbiased estimator of
\begin{eqnarray*}
\frac{(M-1)(B\sigma^{2}_{\text{btw}} + \sigma^{2}_{\text{wtn}}) + M(B-1)\sigma^{2}_{\text{wtn}}}{MB}
\end{eqnarray*}
Schomaker and Heumann \cite{schomaker2018bootstrap} considered large values of $B$ (e.g. 200) and smaller values of $M$. For large $B$ the preceding expression is approximately equal to
\begin{eqnarray*}
(1-M^{-1})\sigma^{2}_{\text{btw}} + \sigma^{2}_{\text{wtn}}
\end{eqnarray*}
Thus if both $M$ and $B$ are large, this is unbiased for $\sigma^{2}_{\text{btw}} + \sigma^{2}_{\text{wtn}}$, which under congeniality is the true posterior variance. If $M$ is not large however, it is biased downwards for the true posterior variance, and so we would expect confidence intervals constructed using the $MB$ sample of estimates, e.g. based on percentiles as suggested by Schomaker and Heumann, to undercover. This concurs with the findings shown in Figure 1 of Schomaker and Heumann, who found that the percentile MI boot pooled confidence intervals undercovered somewhat for small $M$ even under congeniality (Figure 1 of \cite{schomaker2018bootstrap}).

Under uncongeniality, there is no reason to expect this approach to result in valid inferences, and the setting $2$ with high missingness simulation results of Schomaker and Heumann support this, with coverages between 89\% and 92\%.

\subsection{Bootstrap followed by MI}
We now consider methods which first bootstrap sample the observed data and then apply MI to each bootstrap sample. This general approach to combining bootstrap with MI was proposed by Shao and Sitter \cite{shao1996bootstrap} and Little and Rubin \cite{Little/Rubin:2002}.

\subsubsection{Boot MI percentile}
Both Schomaker and Heumann \cite{schomaker2018bootstrap} and Brand \etal \cite{brand2019combining} recommended calculating bootstrap percentile intervals to the estimator $\hat{\theta}_{M}$. This consists of:
\begin{enumerate}
	\item $B$ bootstrap samples of the observed data are taken $Y^{obs}_{b}$, $b=1,..,B$
	\item For each $b=1,...B$, use MI to impute missing data in $Y^{obs}_{b}$ $M$ times, and estimate $\theta$ in each imputed dataset, giving $\hat{\theta}_{b,m}$
	\item For point estimation of $\theta$ one can either use $\hat{\theta}_{M}$ or $\hat{\theta}_{BM}=B^{-1}\sum^{B}_{b=1} \hat{\theta}_{b}$, where $\hat{\theta}_{b} = M^{-1} \sum^{M}_{m=1} \hat{\theta}_{b,m}$.
	\item A $(1-2\alpha)\%$ percentile confidence interval for $\theta$ is then formed by taking the $\alpha$ and $1-\alpha$ empirical percentiles of the $\hat{\theta}_{b}$, $b=1,..,B$ values
\end{enumerate}
This approach to direct application of the standard percentile based bootstrap confidence interval to the estimator $\hat{\theta}_{M}$ and as such, as suggested by Shao and Sitter, we expect it to be asymptotically valid even under uncongeniality \cite{shao1996bootstrap}. Moreover, provided the point estimator is consistent, asymptotically the resulting confidence intervals should attain norminal coverage even if the imputation and/or analysis models are misspecified. In Schomaker and Heumann's setting 2 simulation results, where we believe the imputation and complete data models are uncongenial, they found coverage rates close to 95\%, although for one parameter it was as low as 90\%.

Brand \etal also found that the Boot MI \% approach worked well in simulations \cite{brand2019combining}. They investigated it using either $M=5$ or $M=1$, and among the different combinations of bootstrapping and MI recommended using it $M=1$. Although we expect the resulting confidence intervals to be valid, even under uncongeniality, we expect the intervals to be unnecessarily wide with $M=1$ because only one imputation is used per bootstrap. This is confirmed by the simulation results of Brand \etal \cite{brand2019combining} (Figure 1, panel C), which shows that the bootstrap percentile intervals were wider on average with $M=1$ compared with $M=5$. Moreover, their results suggested that coverage with $M=1$ was slightly above the nominal 95\% level, which we investigate further in Section \ref{refSims}.

\subsubsection{Boot MI von Hippel}
Of the various combinations of bootstrapping and imputation described, assuming the MI point estimator is consistent, only Boot MI percentile is expected to give confidence intervals that attain nominal coverage (asymptotically) under uncongeniality or model misspecification. A practical issue however is that the computational burden is high. For standard applications of MI, it is not uncommon now for $M$ to be chosen as 100 or greater, for reasons of statistical efficiency of point estimates and to reduce Monte-Carlo error to an acceptable amount \cite{White2011, lu2017number, von2018many}. For bootstrap confidence intervals, the number of replications $B$ is generally recommended to be at least 200 for variance estimation and at least 1,000 for percentile based intervals \cite{Efron/Tibshirani:1993}. These considerations would imply a potentially very large value of $BM$, which may be computationally expensive or impractical.

von Hippel and Bartlett proposed an alternative point estimator and confidence interval based on Boot MI \cite{vonHippelbartlett2019}. He proposed using $\hat{\theta}_{BM}=B^{-1}\sum^{B}_{b=1} \hat{\theta}_{b}$ rather than $\hat{\theta}_{M}$, as the point estimator. To construct a confidence interval, von Hippel and Bartlett noted that the estimates $\hat{\theta}_{b,m}$ can be expressed as:
\begin{eqnarray}
\hat{\theta}_{b,m} = \hat{\theta}_{\infty} + c_{b} + d_{bm}
\label{bootmianova}
\end{eqnarray}
where $E(c_{b})=E(d_{bm})=0$, $\Var(c_{b})=\sigma^{2}_{\infty}$, $\Var(d_{bm})=\sigma^{2}_{\text{btw}}$. Given this variance components model, we have that
\begin{eqnarray}
\Var(\hat{\theta}_{BM}) = \left(1+\frac{1}{B} \right) \sigma^{2}_{\infty}  + \frac{1}{BM} \sigma^{2}_{\text{btw}}
\label{vhVar}
\end{eqnarray}
This shows that provided $B$ is large, $\hat{\theta}_{BM}$ will have similar efficiency to $\hat{\theta}_{M}$ with $M$ large. The two variance components $\sigma^{2}_{\infty}$ and $\sigma^{2}_{\text{btw}}$ can be estimated by fitting a one-way analysis of variance to the point estimates $\hat{\theta}_{b,m}$. Letting $MSW$ and $MSB$ denote the mean sum of squares within and between bootstraps, the restricted maximum likelihood estimates of the two variance components are
\begin{eqnarray*}
\hat{\sigma}^{2}_{\infty} &=&  \frac{MSB-MSW}{M}\\
\hat{\sigma}^{2}_{\text{btw}} &=& MSW
\end{eqnarray*}
or if $MSB-MSW<0$, we set $\hat{\sigma}^{2}_{\infty}=0$ and $\hat{\sigma}^{2}_{\text{btw}}$ equal to the total sample variance of the $BM$ estimates. These can be substituted into equation \eqref{vhVar} to estimate the variance of $\hat{\theta}_{BM}$ with
\begin{eqnarray*}
\widehat{\Var}(\hat{\theta}_{BM}) &=& \left(1 + \frac{1}{B} \right)  \frac{MSB - MSW}{M} + \frac{MSW}{BM} \\
&=& \left(\frac{B+1}{BM}\right)  MSB + MSW \left(\frac{1}{BM} - \frac{B+1}{BM} \right) \\
&=& \left(\frac{B+1}{BM}\right)  MSB -\frac{MSW}{M}
\end{eqnarray*}
von Hippel and Bartlett proposed constructing confidence intervals based on Satterthwaite's degrees of freedom, which here is given by
\begin{eqnarray*}
\hat{\nu} &=& \frac{\left[\left(\frac{B+1}{BM}\right)  MSB -\frac{MSW}{M} \right]^{2}}{\frac{\left(\frac{B+1}{BM} \right)^{2}MSB^{2}}{B-1} + \frac{MSW^{2}}{BM^{2}(M-1)} } 
\end{eqnarray*}
If $MSW$ is small (i.e. when the between imputation variance is small), this will be close to $B-1$. A $100 \times (1-\alpha)$ confidence interval for $\theta$ can then be constructed as
\begin{eqnarray*}
\hat{\theta}_{BM} \pm t_{1-\alpha/2,\hat{\nu}} \sqrt{\widehat{\Var}(\hat{\theta}_{BM})}
\end{eqnarray*}
where $t_{1-\alpha/2,\nu}$ denotes the $1-\alpha/2$ quantile of the t-distribution on $\nu$ degrees of freedom. von Hippel and Bartlett advocated use of a large value of $B$ and $M=2$.

\section{Simulations}
\label{simulations}
In this section we report two simulation studies to empirically demonstrate the performance of the previously described combinations of bootstrapping and MI under uncongeniality and/or model misspecification.

\subsection{Regression models under uncongeniality and misspecification}
We first compared the previously described bootstrap and MI combination methods in four scenarios of uncongeniality and/or misspecification of the imputation and analysis models using a simulation study based on one performed by Hughes \textit{et al} \cite{hughes2016}.

Briefly, we simulated a hypothetical dataset of one binary variable, sex, and four continuous variables, age, height, weight and natural log of insulin index (hereafter referred to as loginsindex). The data were generated under the following model:
\begin{equation}\begin{array}{l}
sex \sim Bernoulli(\pi), \; age, height | sex \sim N(\alpha_0 + \alpha_1 sex,\Sigma), \\[4mm]
weight = \iota_0 + \iota_1 sex + \iota_2 age + \iota_3 height + \eta^{sex}\lambda \times error_W, \\[4mm]
loginsindex = \beta_0 + \beta_1 sex + \beta_2 age + \theta weight + \eta^{sex}\omega \times error_L, 
\end{array}\label{data model}\end{equation}
where $error_W$ and $error_L$ are independent errors and $\eta^{sex}=1$ when sex$=0$ and $\eta^{sex}=\eta$ when sex$=1$. Parameter values are shown in Table \ref{HughesParameterValues}. Different scenarios were created by setting parameters $\alpha_1, \iota_1, \nu$ and $\beta_1$ to their null values. The values of the remaining parameters were fixed. Weight measurements were set to be missing completely at random for $60\%$ of the observations. 

The analysis of interest was to estimate $\theta$, the effect of weight on loginsindex after adjustment for age and sex. Both imputation and analysis models were normal linear regression models that assumed homoscedastic errors. Unless otherwise stated, the distributions of $error_W$ and $error_L$ were normal, weight measurements were missing in men and women, the assumption of homoscedastic errors was true, and the imputation and analysis models were fitted to the entire sample. The following four scenarios were considered:
\begin{description}
	\item[Subgroup analysis scenario] The data were simulated such that the continuous variables were identically distributed in men and women, and weights was missing among men only. The imputation and analysis models were uncongenial since the analysis model was fitted to men only whilst the imputation model was fitted to the entire sample ignoring sex (i.e. excluding sex as a predictor). 

	\item[Heteroscedastic errors] The data were simulated such that the variance of weight and loginsindex differed between men and women. The imputation and analysis models were congenial but incorrectly specified because they assumed homoscedastic errors.
	
	\item[Omitted interaction] As in all scenarios, the data were simulated such that the effect of weight on loginsindex was the same for men and women. The imputation and analysis models were uncongenial because the analysis model included an interaction term between weight and sex whilst this interaction was, correctly, omitted from the imputation model. 
	
	\item[Moderate non-normality] Error distributions $error_W$ and $error_L$ were simulated from the log-normal distribution $\exp\{N(0,1/4^2)\}$. The imputation and analysis models were congenial, but misspecified because they assumed a normal error distribution.
\end{description}

For each scenario we generated $1,000$ independent simulated datasets, where the sample size was $1,000$ observations and the probability of observing weight was $0.4$, except for the subgroup analysis scenario where the probability of observing weight was $1$ among women and $0.4$ among men. We conducted MI Rubin using $10$ imputations, and methods MI boot Rubin, MI boot pooled percentile and boot MI percentile with $10$ imputations and $200$ bootstraps, and von Hippel's boot MI with $2$ imputations and $200$ bootstraps. Additionally, we applied boot MI percentile with $1$ imputation and $200$ bootstraps. Based on $1,000$ simulations the Monte Carlo standard error for the true coverage probability of $95\%$ is $\surd(0.95(1-0.95)/1000)=0.69\%$, implying that the estimated coverage probability should lie within the range $0.936$ and $0.964$ (with $95\%$ probability) \cite{morris2019using}. 

For all methods, the point estimates of $\theta$ were either unbiased or the amount of systematic bias was trivial (e.g., at most $-0.000289$; results available on request). 

Tables \ref{rachMainResults} and \ref{rachTable2} show the median of the CI widths and CI coverage for the $6$ methods under comparison. For the subgroup analysis scenario (Table \ref{rachMainResults}), MI Rubin and both MI then bootstrapping methods resulted in confidence interval overcoverage. Narrower confidence intervals and nominal coverage were achieved with the boot MI percentile method with $10$ imputations and boot MI von Hippel. Boot MI percentile with single imputation resulted in wide confidence intervals and overcoverage. This concurs with what was found in the simulations reported by Brand \textit{et al}. In the Appendix we give a sketch argument for why the Boot MI percentile intervals with $M=1$ (or indeed small $M$ more generally) will overcover. Interestingly, this over-coverage does not similarly affect normal based (as opposed to percentile) Boot MI intervals with $M=1$.

For the heteroscedastic errors scenario (Table \ref{rachMainResults}), MI Rubin and both MI then bootstrapping methods resulted in confidence interval undercoverage. Again, the boot MI percentile method with $10$ imputations and boot MI von Hippel were the best performing methods with close to nominal coverage.  The results for the omitted interaction scenario (Table \ref{rachTable2}) followed a similar pattern noted for the subgroup analysis scenario. For the moderate normality scenario (Table \ref{rachTable2}), MI boot pooled percentile had slight confidence interval undercoverage and boot MI percentile with single imputation overcovered. The remaining methods had close to nominal coverage with similar median CI widths. 

\begin{table}[ht]
\small\centering
\scalebox{1}{
\begin{threeparttable}
\parbox{25cm}{\caption{\scriptsize{The values of the data model parameters to four significant figures.}}}\\
{\scriptsize\addtolength{\tabcolsep}{-2pt}
\begin{tabular}{l c}
\hline \\[-8pt]
Parameter  & Value(s) \\ 
\hline \\[-8pt]
$\pi$ & $0.4577$ \\[2mm]
$\alpha_0$ & $(25.02, 1.774)$ \\[2mm]
$\alpha_1$ & $(-0.03616, -0.1336)$ \\[2mm]
$\Sigma$ & $(0.5521, 0.001574$ \\ 
         & $0.001574, 0.003705)$ \\[2mm]
$\iota_0$ & $-32.98$ \\[2mm]
$\iota_1$ & $-2.314$ \\[2mm]
$\iota_2$ & $-0.01566$ \\[2mm]
$\iota_3$ & $65.38$ \\[2mm]
$\lambda$ & $12.29$ \\[2mm]
$\beta_0$ & $1.854$ \\[2mm]
$\beta_1$ & $0.2908$ \\[2mm]
$\beta_2$ & $0.08003$ \\[2mm]
$\beta_3$ & $0.01119$ \\[2mm]
$\omega$  & $0.7887$  \\[2mm]
$\eta$    & $0.5$ \\
\hline
\end{tabular}}
\label{HughesParameterValues}
\end{threeparttable}}
\end{table}

\begin{table}[ht]
\caption{Median confidence interval width and coverage for the subgroup analysis (uncongenial) and heteroscedastic errors (misspecification) scenarios.}
\label{rachMainResults}
\centering
\begin{tabular}{lllllll}
 \toprule
 & & &  \multicolumn{2}{c}{Subgroup} & \multicolumn{2}{c}{Heteroscedastic} \\
 & & &  \multicolumn{2}{c}{analysis} & \multicolumn{2}{c}{errors} \\
 \midrule
 &  & &  Median  &  & Median  & \\ 
  & $M$ & $B$ & CI width & CI cov. &  CI width & CI cov. \\ 
  \midrule
MI Rubin &10&   & $0.0142$ & $98.2$ & $0.0126$ & $91.3$ \\ 
  MI boot Rubin & 10 & 200  & $0.0143$ & $98.1$ & $0.0129$ & $92.1$ \\ 
  MI boot pooled percentile & 10 & 200 & $0.0131$ & $97.7$ & $0.0117$ & $89.2$ \\ 
  Boot MI percentile & 10 & 200 & $0.0109$ & $94.9$ & $0.0144$ & $95.0$ \\ 
  Boot MI percentile & 1 & 200 & $0.0139$ & $98.4$ & $0.0167$ & $97.7$ \\ 
  von Hippel & 2 & 200 & $0.0108$ & $95.0$ & $0.0144$ & $94.1$ \\ 
   \bottomrule
\end{tabular}
\end{table}

\begin{table}[ht]
\caption{Median confidence interval width and coverage for the omitted interaction (uncongenial) and moderate non-normality (misspecification) scenarios.}
\label{rachTable2}
\centering
\begin{tabular}{lllllll}
 \toprule
 & & &  \multicolumn{2}{c}{Omitted } & \multicolumn{2}{c}{Moderate} \\
 & & &  \multicolumn{2}{c}{interaction} & \multicolumn{2}{c}{non-normality } \\
 \midrule
 &  & &  Median  &  & Median  & \\ 
  & $M$ & $B$ & CI width & CI cov. &  CI width & CI cov. \\ 
  \midrule
MI Rubin &10&   & $0.0146$ & $97.3$ & $0.0119$ & $94.6$ \\ 
  MI boot Rubin & 10 & 200  & $0.0146$ & $97.2$ & $0.0120$ & $94.7$ \\ 
  MI boot pooled percentile & 10 & 200 & $0.0135$ & $95.4$ & $0.0108$ & $93.1$ \\
  Boot MI percentile & 10 & 200& $0.0128$ & $94.2$ & $0.0118$ & $95.4$ \\ 
  Boot MI percentile & 1 & 200 & $0.0159$ & $98.0$ & $0.0143$ & $98.1$ \\ 
  von Hippel & 2 & 200 & $0.0127$ & $94.0$ & $0.0117$ & $95.1$ \\ 
   \bottomrule
\end{tabular}
\end{table}

\subsection{Reference based imputation in clinical trials}
\label{refSims}
Our second simulation study setting is so called control or reference based MI for missing data in randomised trials. Missing data due to study dropout is common on clinical trials, and there is often concern that missing data do not satisfy the missing at random (MAR) assumption. Often dropout in trials coincides with patients' treatments changing. An increasingly popular approach to imputing missing data in trials is using so called reference or control based MI approaches \cite{Carpenter2013}. These involve constructing the imputation distribution for the active treatment arm using a combination of information from the active and control arms, which results in uncongeniality between imputation and analysis models. This uncongeniality results in intervals constructed using Rubin's variance estimator to over-cover \cite{seaman2014comment,tang2017}. Cro \textit{et al} have suggested that although Rubin's variance estimator is biased for the repeated sampling variance of the estimator, it consistently estimates a sensible variance in the context of MAR sensitivity analyses \cite{cro2019information}. We do not here enter this debate, but merely investigate the previously described bootstrap and MI combinations in regards their ability to produce confidence intervals with the correct repeated sampling coverage. In the setting of reference based MI Quan \textit{et al} applied (we believe) Boot MI to estimate standard errors of $\hat{\theta}_{M}$, and found it worked well \cite{quan2018considerations}.

We simulated 10,000 datasets of size $n=500$ with 250 randomised to control ($Z=0$) and 250 ($Z=1$) randomised to active treatment. Baseline $X$ and outcome $Y$ were then generated from a bivariate normal model:
\begin{eqnarray*}
\begin{pmatrix} X \\ Y \end{pmatrix} \sim N \left(\begin{pmatrix} 2 \\ 2+0.2Z \end{pmatrix}, \begin{pmatrix} 0.4 & 0.2 \\ 0.2 & 0.4 \end{pmatrix} \right) 
\end{eqnarray*}
The analysis model was normal linear regression of $Y$ on $X$ and $Z$, with the coefficient of treatment $Z$ of primary interest. Values in $Y$ were made missing complete at random with probability 0.5. For each dataset, first the missing values in $Y$ were imputed using a normal linear regression model with $X$ and $Z$ as covariates assuming MAR, such that the imputation and analysis models were congenial. Second they were imputed using the jump to reference method (see \cite{Carpenter2013} for details), such that the two models were uncongenial. The same combinations of bootstrapping and MI were used as in the first simulation study.

Table \ref{j2rResults} shows the median confidence interval width and coverage for each of the combinations of bootstrapping and MI previously described. Based on $10,000$ simulations the Monte Carlo standard error for the true coverage probability of $95\%$ is $\surd(0.95(1-0.95)/10000)=0.43\%$. As expected, intervals constructed using Rubin's rules have correct coverage under congeniality. Under jump to reference imputation, where the imputer assumes more than the analyst \cite{seaman2014comment}, Rubin's variance estimator is biased upwards and intervals over-cover. Intervals constructed using MI boot Rubin perform well under MAR (congeniality) but like standard Rubin's rules over-cover under uncongeniality as anticipated. MI boot pooled percentile under-covers somewhat under congeniality, which following the earlier explanation is due to the relatively small choice of $M$. Under uncongeniality these intervals over-cover, since again their justification relies on congeniality.

Both Boot MI percentile and the Boot MI von Hippel approach with $B=200$ and $M=2$ give intervals with approximately correct coverage under both congeniality and uncongeniality, but the von Hippel intervals are computationally much quicker. As in the first simulation study, the Boot MI percentile intervals with $B=200$ and $M=1$ over-covered, even under congeniality.

\begin{table}[ht]
\caption{Median confidence interval width and coverage under MAR (congenial) and jump to reference (uncongenial) imputation from 10,000 simulations.}
\label{j2rResults}
\centering
\begin{tabular}{lllllll}
 \toprule
 & & &  \multicolumn{2}{c}{MAR } & \multicolumn{2}{c}{Jump to reference } \\
 & & &  \multicolumn{2}{c}{(congenial)} & \multicolumn{2}{c}{(uncongenial) } \\
 \midrule
 &  & &  Median  &  & Median  & \\ 
  & $M$ & $B$ & CI width & CI cov. &  CI width & CI cov. \\ 
  \midrule
MI Rubin &10&   & 0.285 & 94.09 & 0.251 & 99.73 \\ 
  MI boot Rubin & 10 & 200  & 0.284 & 94.11 & 0.251 & 99.70 \\ 
  MI boot pooled percentile & 10 & 200 & 0.258 & 91.79 & 0.236 & 99.53 \\ 
  Boot MI percentile & 10 & 200 & 0.272 & 94.24 & 0.154 & 94.89 \\ 
  Boot MI percentile & 1 & 200 &  0.326 & 97.55 & 0.207 & 99.21 \\ 
  von Hippel & 2 & 200 & 0.275 & 94.26 & 0.153 & 94.80 \\ 
   \bottomrule
\end{tabular}
\end{table}

\section{Discussion}
\label{discussion}
We have reviewed a number of proposals for combining MI with bootstrapping, in particular in regards their statistical validity when imputation and analysis models are uncongenial or misspecified. Approaches which first impute then bootstrap generally do not give valid inferences under uncongeniality or model misspecification. In contrast, bootstrapping followed by imputation is robust to uncongeniality, and provided the MI point estimator is consistent, the resulting confidence intervals have the correct coverage asymptotically even if the imputation and/or analysis models are misspecified. A drawback of this approach with $M$ large is its high computational cost. Brand \etal recommended this method, but using $M=1$ imputation per bootstrap, which obviously reduces the computational burden considerably \cite{brand2019combining}. However, with small $M$, the MI point estimator is inefficient, and moreover we have shown that percentile based Boot MI intervals over-cover even under congeniality for small $M$. The alternative boot MI version proposed by von Hippel and Bartlett overcomes these drawbacks, only requiring $BM$ imputations and analyses, and $M$ can be chosen to be two. It does however, like Rubin's rules, assume that the MI estimator is normally distributed. The Boot MI von Hippel approach is implemented in the R package bootImpute, and is available from CRAN \cite{bootImpute}.  As far as we are aware the only alternative approaches for valid inferences under uncongeniality require complex problem specific calculations which are not conducive to general use \cite{Robins/Wang:2000,tang2017}, and in this context the Boot MI von Hippel approach seems very attractive.

As mentioned in the introduction, Rubin originally envisaged the imputer and analyst as distinct entities, with the imputer releasing a single set of multiply imputed datasets to different analysts. A strength of the bootstrap followed by MI approach is that this division of roles is still feasible - the imputer bootstraps and then multiply imputes the observed data, releasing a set of imputations clustered by bootstrap. These can then be analysed by different analysts, and inferences can be obtained using either the boot MI percentile or Boot MI von Hippel approaches.

Combining bootstrapping with MI has some disadvantages compared to inference using Rubin's rules. Compared to regular MI with Rubin's rules, it is considerably more computationally intensive - this is the price paid for being able (in certain situations) to obtain valid inferences under uncongeniality or misspecification. Problems with model (imputation or analysis) convergence are probably more likely due to the large number of bootstraps required. The non-parametric resampling scheme used by bootstrapping relies on an assumption that the data are independent and identically distributed, and further research is warranted to explore the use of other types of bootstrap resampling schemes in conjunction with MI.

Code for the first simulation study (R) and the second simulation study (Stata) are available from https://github.com/jwb133/bootImputePaper.

\section*{Acknowledgements}
This research made use of the Balena High Performance Computing (HPC) Service at the University of Bath.

\section*{Funding}
RAH was supported by the Medical Research Council Integrative Epidemiology Unit at the University of Bristol (MC\_UU\_00011/3) and a Sir Henry Dale Fellowship jointly funded by the Wellcome Trust and the Royal Society (Grant Number 215408/Z/19/Z).

\section*{Appendix}
In this appendix we sketch an argument for why percentile based bootstrap confidence intervals over-cover with small values of $M$. To simplify the argument, we will assume that the bootstrap distributions of estimates are normal. Consider first Boot MI with $M=\infty$. Following equation \eqref{bootmianova}, and assuming the bootstrap distribution is normal, this distribution will be $N(\hat{\theta}_{\infty}, \sigma^{2}_{\infty})$. If the bootstrap distribution is normal, the percentile interval is equal (with $B=\infty$) to $\hat{\theta}_{\infty} \pm 1.96 \sigma_{\infty}$. Now suppose that this confidence interval, in repeated samples, has correct coverage.

Now consider the same procedure with small finite $M$. Following equation \eqref{bootmianova}, the bootstrap distribution of estimates is now $N\left(\hat{\theta}_{\infty}, \sigma^{2}_{\infty}+\frac{\sigma^{2}_{\text{btw}}}{M}\right)$. The resulting boot MI percentile confidence interval (with $B=\infty$) is then $\hat{\theta}_{\infty} \pm 1.96 \sqrt{ \sigma^{2}_{\infty}+\frac{\sigma^{2}_{\text{btw}}}{M}}$. The lower limit of this interval is then less than the lower limit of the interval with $M=\infty$, and the upper limit is larger than the upper limit of the interval with $M=\infty$. Hence if the interval with $M=\infty$ has correct coverage, when $M$ is finite, the percentile interval with $M$ finite must over-cover. Note that this argument does not apply to a normal based Boot MI interval, because this interval is constructed as $\hat{\theta}_{M} \pm 1.96 \sqrt{ \sigma^{2}_{\infty}+\frac{\sigma^{2}_{\text{btw}}}{M}}$.

\end{document}